\documentstyle[12pt]{article}
\newcommand {\PRep}  {Phys.\ Rep.}

\newcommand {\ZP}    {Z.\ Phys.}
\newcommand {\NP}    {Nucl.\ Phys.}
\newcommand {\PL}    {Phys.\ Lett.}
\newcommand {\T}    {\perp}
\newcommand {\pT}   {p_\T}

\newcommand {\hT}   {h_\T}
\newcommand {\kT}   {k_\T}
\newcommand {\Pol}  {{\cal P}}
\newcommand {\vpT}  {\vec p_\T}
\newcommand {\vqT}  {\vec q_\T}
\newcommand {\vkT}  {\vec k_\T}
\newcommand {\vhT}  {\vec h_\T}

\newcommand {\vPol} {\vec{\cal P}}
\newcommand {\Aqq}  {{\cal A}_{q \to q'}}
\newcommand {\up}   {\!\!\uparrow\!\!}

\newcommand {\beq}  {\begin{equation}}
\newcommand {\eeq}  {\end{equation}}

\title{Single spin asymmetry of vector meson production \\ 
       in the string model}
\author{{\sc J.\,Czy\.zewski\thanks{
Fellow of the Polish Science
Foundation (FNP) scholarship for the year 1996 }}\\
{\it Institute of Physics, Jagellonian University,} \\
{\it ul.\,Reymonta 4, PL-30\,059 Krak\'ow, Poland.}
\vspace{0.5cm} \\
{\normalsize\sf Dedicated to Professor
Andrzej Bia\l{}as in honour of his 60th birthday}}
\date{}
\begin{document}
\begin{titlepage}
\pagestyle{empty}
\begin{center}
\maketitle
\end{center}
\begin{abstract}
Azimuthal asymmetry of vector-meson production in single-transversely polarized
proton-proton collisions ($p\up p$) is calculated in a model based
on the string fragmentation. The string is spanned by a valence quark of
the projectile scattered on the target.
The asymmetry is generated only during fragmentation of the scattered
quark into hadrons.  
The obtained asymmetry of the $\rho$ mesons is
opposite in sign to that of pions.  On the other hand, if the asymmetry
were generated during the quark scattering then the asymmetries of the vector
and of the pseudoscalar mesons would be close to each other.

\end{abstract}

{\large\sf TPJU 11/96}

{\large\sf May 1996}

\end{titlepage}
\newpage
Measurement of single transverse-spin azimuthal asymmetries of particle
production in $p\up p$ collisions can provide information about transversely
polarized quark distributions in the proton \cite{Collins-PANIC}. However, 
to extract
information about the latter from the experimental data, one has to know the
mechanism generating the asymmetry.
It has been shown recently \cite{ACY} that one can obtain large asymmetries
with their signs and $x_F$ dependence 
similar to those measured by the E704 experiment \cite{E704-91} while
assuming that the asymmetry is generated only during the fragmentation 
of polarized quarks produced in the collision. Such an effect was originally
proposed by Collins \cite{Collins93}.
In Ref.~\cite{ACY} the string model was used to
describe the fragmentation, and the polarization effects were parametrized as
prescribed by the Lund model \cite{LUND}.  Positive asymmetry was obtained for $\pi^+$
and $\pi^0$ production and negative for $\pi^-$, resulting from upward
(downward) polarizations of the $u$ ($d$) valence qark in the proton polarized
upwards.

However, different mechanisms leading to the azimuthal asymmetry are possible.
Szwed has shown \cite{Szwed81,Szwed90} that the asymmetry appears in 
scattering of a quark on an external strong field due to multiple gluon 
exchange.  This asymmetry
vanishes at a sufficiently high energy due to chiral symmetry and the resulting
helicity conservation.  Nevertheless, it can be measurable at finite energies.

In this note we show that comparing the asymmetries of vector and pseudoscalar
mesons can provide information on the magnitude of the asymmetry of quark
scattering. We calculate the asymmetry of $\rho$ mesons along the lines of
Ref.~\cite{ACY}.  This asymmetry is opposite in sign to that of pions. 
On the contrary, if the asymmetry of the quark scattering were  the dominating
one, then the asymmetries of pseudoscalar
(PS) and vector (V) mesons would not differ much.

We consider the reaction:
\beq 
p\!\uparrow + \, p \rightarrow h + X,
\label{reaction}
\eeq
where $\uparrow$ refers to the projectile proton polarized vertically upwards
(parallel to the $\hat y$ axis) if the beam momentum points in the $\hat z$
direction.  The produced hadron $h$ carries the fraction $x_F = 2 p_z /
\sqrt{s}$ of the center-of-mass longitudinal momentum and the transverse
momentum $\vec \pT$.  In measurements of the asymmetry, the polarized
cross-section $d\sigma_\uparrow$ is assumed to behave as

\beq
d\sigma_\uparrow(x_F, \vec\pT)= d\sigma(x_F,\pT)
   [1 + A_N(x_F,\pT) \cos(\phi)],
\label{eq2}
\eeq
where $d\sigma$ denotes the unpolarized cross-section.  This defines the
asymmetry $A_N$.  $\phi$ is the azimuthal angle of the transverse momentum
$\vec\pT$ of the hadron, measured with respect to the $\hat x$ axis.

If one assumes factorization, the cross-section for the reaction 
(\ref{reaction}) is a convolution of the parton distribution $q(x)$,
the cross-section $d\hat\sigma$ for the parton scattering and the
fragmentation function $D_{h/q}(z)$ of the scattered quark $q$ into
the hadron $h$.

When the projectile proton is transversely polarized then the quark
acquires also transverse polarization $\vPol_q$, its magnitude 
being defined by the transversity distribution \cite{Ralston79}
\beq
\Pol_q(x) = {\Delta_\T q(x) \over q(x)}
\eeq
depending on the quark momentum fraction $x$.
After scattering this polarization can be diminished by a depolarization
factor $D_{NN}$. However, $D_{NN}$ is close to unity at typical small
scattering angles \cite{Collins93} and the polarization is conserved 
during scattering. Moreover, the transverse momentum $\vec\pT$ of the 
produced hadron is the sum of its part $z\vqT$ inherited from the 
transverse momentum $\vqT$ of the scattered quark and some additional 
transverse momentum $\vhT$ coming from the fragmentation.

Hence, the spin effects can be included by the dependence of 
the fragmentation function $D_{h/q}$ on
the quark polarization $\vPol_{q}$ (Collins effect).
They manifest themselves in an asymmetry of production of the hadrons
in the azimuth of
the transverse momentum $\vec h_\T$ of the hadron with respect to the
axis of the scattered quark.  

Another source of asymmetry can appear
at the parton level {\it i.e.}  when $d\hat
\sigma$ depends on the azimuth $\hat\varphi$ of the transverse momentum
$\vqT$ of the scattered quark (Szwed effect \cite{Szwed81,Szwed90}).

Hypothetically, two extreme cases could be defined:

\begin{itemize}
\item[\it a)] the asymmetry appears only in the fragmentation function as
calculated in \cite{ACY}. In this case it can depend on whether the produced
hadron is a PS or V meson. 
This is the case of the high-energy limit since 
at a sufficiently high energy any asymmetry appearing at the parton level 
(in $d\hat \sigma$) must vanish due to chiral symmetry.  

\item[\it b)] The asymmetry appears only at the parton scattering.
It has been shown by Szwed in \cite{Szwed90} that this mechanism can lead to 
significant asymmetries at the beam energy of the order of
$10-20\,$GeV.  Here, the final asymmetry is defined before fragmentation 
and it does not depend on whether $h$ is a PS or a V meson.
It can depend, however, on the flavour of the scattered quark.
\end{itemize}

\noindent
In reality, one can expect a mixture of the two effects with the second one
vanishing at a sufficiently high energy.  We shall concentrate on the case
{\it a)} and calculate the asymmetry of vector mesons therein.

In reference \cite{ACY} it was argued that the polarized cross-section
for the reaction (\ref{reaction}) at large positive $x_F$ can be written as:
$$
{d\sigma \over dx_F d^2\vpT} =
\sum_{q=u,d}\int dx q(x) \int d^2\vqT {d\hat\sigma \over d^2 \vqT}
\int dz d^2\vhT D_{h/q}(z,\vhT) \times
$$
\beq
\delta\left(x_F - \sqrt{z^2x^2-{4\pT^2\over s}}\right)
\delta^2(\vpT - z \vqT - \vhT).
\label{xsection}
\eeq
The fragmentation function $D_{h/q}$ was calculated there in the framework
of the string model \cite{LUND} of particle production. In order to
include the spin effects, it was divided into two parts:
\beq
D_{h/q} = D^{\rm rank=1}_{h/q} + D^{{\rm rank}\, \ge\, 2}_{h/q}
\eeq
where $D^{\rm rank=1}_{h/q}$ corresponds to the first-rank (leading)
hadron containing the original fragmenting quark. Only this part
is azimuthally asymmetric.

Two factors determine the spin asymmetry of fragmentation:
\begin{itemize}
\item[a)]
Correlation of the polarization of the quark and the antiquark
$\vPol_q=\vPol_{\bar q}$ of a pair produced in the string and their
orbital angular momentum $L=2k_\T^2 /\kappa$, where $\kT$ and $-\kT$ are the
transverse momenta of $q$ and $\bar q$ and $\kappa$ is the string tension.
This correlation is parametrized in the Lund model \cite{LUND} according
to the formula: $\Pol_q = L/(\beta + L)$ with $\beta$ being a parameter
determined to be between 1 and 2. In this paper, following Ref.~\cite{ACY},
we use $\beta = 1$.
\item[b)]
Probability that the leading quark having the polarization $\Pol_q$ and
the first subleading one with polarization $\Pol_{\bar q}$ form a particular
meson. This probability for the PS mesons (in the nonrelativistic quark
model) is
\beq
{1 \over 4} \, (1 - \vPol_{q} \cdot \vPol_{\bar q}) \,,
\eeq
while for the V mesons it is
\beq
{1 \over 4} \, (3 + \vPol_{q} \cdot \vPol_{\bar q}).
\eeq
This makes the asymmetry of the fragmentation into the 
$\rho$ mesons opposite in sign and 3 times smaller than for pions.
\end{itemize}

In the numerical calculations we used the same parameters as those in
Ref.~\cite{ACY} in which one can find the detailed description of the
model. The only difference between the present calculation an that of
Ref.~\cite{ACY} is in the splitting function determinig
the fragmentation of the string. The heavier resonances have a harder spectrum.
For the $\rho$ mesons, we used the Standard Lund splitting function
$f(z)= (1+C)(1-z)^C$ with the parameter $C=-0.2$,
while for the pions $C=0.5$ was used. This accounts reasonably well for 
the measured inclusive spectra \cite{Aguillar91,Akesson82} 
of the $\rho$ mesons as shown in Fig.~1. The $x_F$ dependence of the
inclusive cross-section is very well reproduced by the model. The obtained 
$p_T$ spectrum is slightly steeper than the experimental one but the 
agreement is satisfactory in view of the simplicity of the model. 
One could probably get a better agreement using the symmetric splitting 
function instead of the Standard-Lund one quoted above.
Unfortunately one cannot get analytical results neither for the spectra 
nor for the asymmetry with the symmetric splitting function.

In Figs~2 and 3 we show the results obtained for the asymmetry $A_N$ of the $\rho$
meson production, at $200\,$GeV beam momentum, compared to that of the charged
pions. 
As already argued, the ratio of the asymmetries is

\beq
R_{\rho/\pi} = {A_N^\rho \over A_N^\pi} \approx -{1\over 3}
\label{ratio}
\eeq
It would be of great interest to verify this prediction experimentally.
The formula (\ref{ratio}) is exact at the level of the fragmentation 
but the slight difference in between $R_{\rho/\pi}$ and $-1/3$ is caused
by the different parameter $C$ used for fragmentation into the pions
and into the $\rho$ mesons.

Violating the rule (\ref{ratio}) can be an indication of appearing of the 
Szwed effect
\cite{Szwed90,Szwed81} {\it i.e.\ }significant asymmetry in parton scattering.  In that model,
originally constructed to describe the polarization of hyperons $\Lambda$ 
\cite{Szwed81},
the asymmetry of a transversely polarized quark scattered on a Coulomb-like
strong field is given by:

\beq 
\Aqq = 2 C_S \alpha_S
       {m q \sin^3(\theta/2) \ln[\sin(\theta/2)]
       \over
       [m^2 + q^2 \cos^2(\theta/2)] \cos(\theta/2) }
       {\vec q \times \vec q' \over |\vec q \times \vec q' |}
\label{qasym}
\eeq
where $m$, $\vec q$ and $\vec q'$ are the mass, the initial and the final
momentum of the scattered quark and $q = |\vec q|$.  $\theta$ is the scattering
angle in the frame where $|\vec q| = |\vec q'|$.  $C_S$ is the constant
characterizing the external strong field.  The sign of the asymmetry depends on
the sign of the constant $C_S$ in (\ref{qasym}) or on whether the field source 
is ``quark-like'' or ``antiquark-like''.  In the first case $C_S$ is positive 
and the asymmetry $\Aqq$ negative, in the latter $C_S$ is negative and $\Aqq$
positive.  It would be interesting to check what the asymmetry is in 
quark--gluon
scattering. 

Hence, if the asymmetry of quark scattering is treated as a correction to that
of fragmentation, $R_{\rho/\pi} < -{1\over 3}$ will mean negative $\Aqq$ or
scattering off a ``quark-like'' field and $-{1\over 3} < R_{\rho/\pi}$ will
mean positive $\Aqq$ and ``antiquark-like'' field.  As explicitly seen from
Eq.~(\ref{qasym}), at sufficiently high energy ($k\gg m$) the asymmetry of quark
scattering vanishes.  The energy scale where it appears can be an interesting
hint about the scale of the masses of partons being scattered in a $pp$
collision.

$R_{\rho/\pi}\ne -{1\over 3}$ could also mean a violation of the 
nonrelativistic
quark model, which was assumed in the calculation of the asymmetry. If 
this were the
case, then the high-energy limit of $R_{\rho/\pi}$ could be taken as the
reference value, instead of $-{1\over 3}$ and the above analysis could be also
made.

To summarize, we have calculated the single spin asymmetry of $\rho$ meson
production in $p\up p$ collisions in the framework of Ref.~\cite{ACY}, 
where the asymmetry of pions has been obtained.  These results are valid for
large positive $x_F$. The asymmetry was generated only in the
fragmentation function using the string model of particle production.  
The two asymmetries are found to be opposite in sign,
the ratio of that of $\rho$ to that of $\pi$ being approximately -1/3.  
Violating this rule
can indicate to a significant contribution from the azimuthal asymmetry of 
the transversely polarized parton subprocess.

This work could not have been completed without the knowledge of the string
model which I was taught by Professor A.~Bia\l{}as. I would like to 
express my gratitude for all his help, patience and understanding.
I am grateful to Professor X.~Artru for collaboration and many enlightening
discussions.
Some claryfying remarks from Professor J.~Szwed are
also acknowledged.

This work has been partially supproted by the Polish Government grant of KBN
no.~2~P03B~083~08 and by the Polish-German Collaboration Foundation grant
FWPN no.~1441/LN/94. Support from the IN2P3-Poland collaboration no.\ 91-62
is also acknowldged.

%
\newpage
\noindent {\Large \bf Figure Captions}
\begin{itemize}
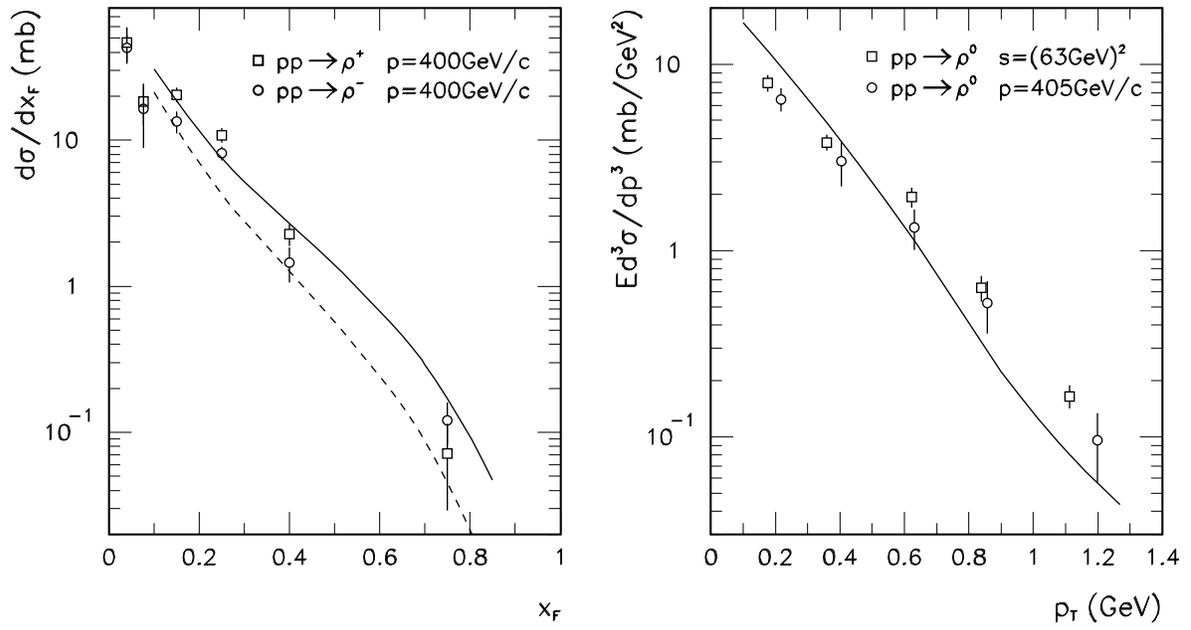
\item[Fig.~1] 
Inclusive cross sections for $\rho$ production as calculated  according to the
formula (\ref{xsection}) compared to data of Ref.~\cite{Aguillar91} ($x_F$ dependence)
and Ref.~\cite{Akesson82} ($\pT$ spectrum).
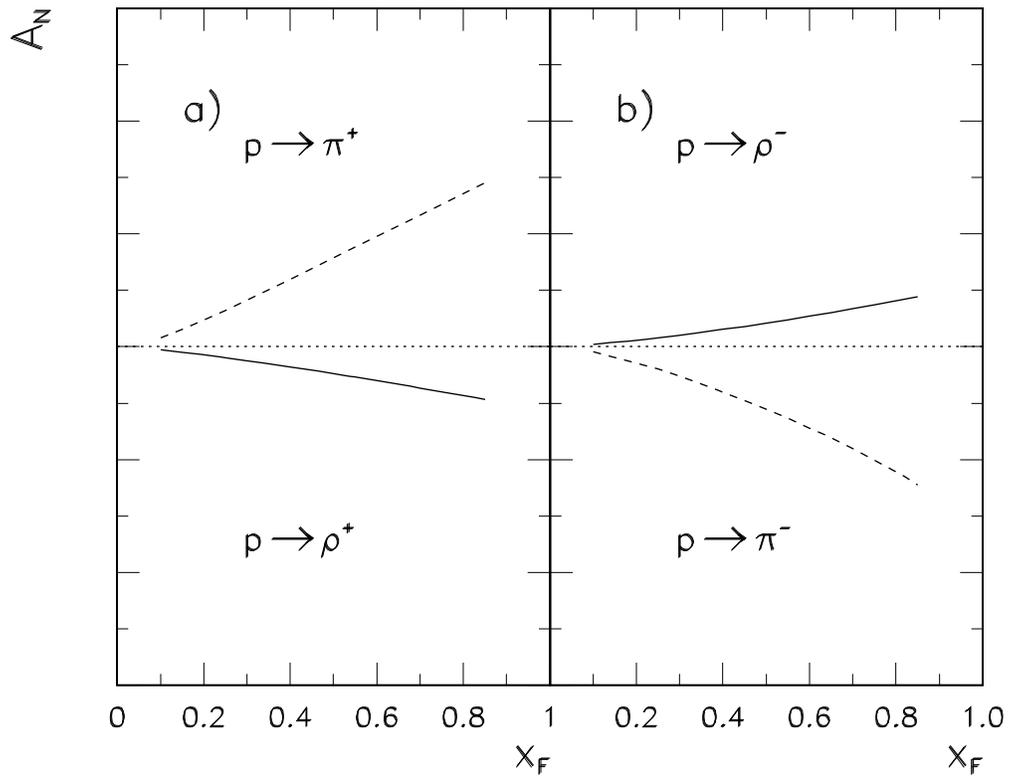
\item[Fig.~2]
The single-spin asymmetry of the charged $\rho$ mesons (full lines) and the 
charged $\pi$ mesons
(dashed lines) plotted versus $x_F$. For all the plots the beam energy of
200GeV was taken and the particle yields used to calculate the asymmetry was 
integrated over $0.5<\pT<2$GeV.
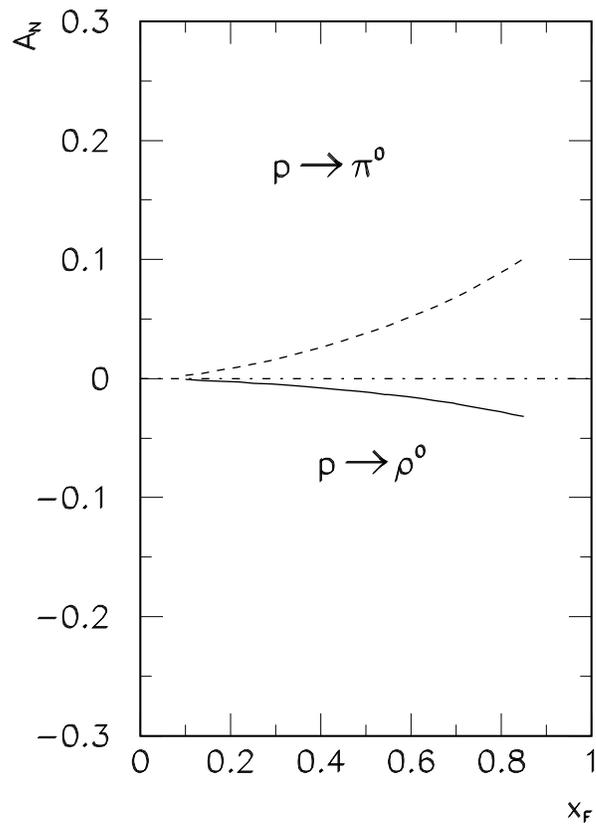
\item[Fig.~3]
The single-spin asymmetry of the neutral $\rho$ mesons (full line) and the 
neutral $\pi$ mesons (dashed line). All parameters are as in Fig.~2.
\end{itemize}
\end{document}